\documentstyle[aps,preprint]{revtex}
\begin{document}
\draft
\title{
Bifurcations in Globally Coupled Chaotic Maps}

\author{
Satoru Morita\thanks{E-mail address: morita@ton.scphys.kyoto-u.ac.jp.}
}

\address{
Department of Physics, Graduate School of Sciences,\\
Kyoto University, Kyoto 606, Japan
}

\maketitle

\begin{abstract}

We propose a new method to investigate collective behavior in a network of
globally coupled chaotic elements generated by  a tent map.
In the limit of large system size, the dynamics
is described with the nonlinear Frobenius-Perron equation.
This equation can be transformed into a simple form by making use of the
piecewise linear nature of the individual map.
Our method is applied successfully to the analyses of stability
of collective stationary states and their bifurcations.
\end{abstract}



\section{Introduction}
Globally coupled dynamical systems form an important class of models in
nonlinear
dynamics. The all-to-all nature of the coupling may be regarded
as an idealization of long-range coupling or an approximation to a short-range
coupling high-dimensional lattice.
Such systems are useful in modeling
diverse phenomena such as Josephson junction arrays\cite{hadley}, multimode
laser\cite{wiesenfeld}, charge-density wave\cite{strogatz},
evolutionary dynamics\cite{ikegami}, biological information
processing and neurodynamics\cite{somplinsky,golomb,kaneko1994,nozawa}.
A simplifying assumption is often made that the coupling strength is uniform
over all coupled pairs.
This makes the theoretical treatment much easier compared with systems
with short-range interaction\cite{kuramoto}.

In the present paper, we study globally coupled maps (GCM) introduced by Kaneko
\cite{kaneko1990a}. They form a dynamical system of $N$
local mappings which are under a common internal field.
The system is thus a mean-field version of coupled map lattices (CML)
\cite{kaneko1989,chate}.
The explicit form of GCM we will work with is given by
\begin{equation}
x_{n+1}(i)=(1-\epsilon)f(x_{n}(i))+{\epsilon\over N}\sum_{i'=1}^Nf(x_{n}(i')) .
\label{maps-defn}
\end{equation}
Here $n$ represents a discrete time step, $i$ the index of the elements
($i=1,2,3,...,N$), and $\epsilon$ the coupling constant.
It is assumed that the map $f(x)$ gives rise to chaotic dynamics.
We specifically consider a tent map
\begin{equation}
f(x)=1-a|x|
\end{equation}
with parameter $a$.
The piesewise linearity of the map simplifies the analysis considerably as we
see later.
The range of $x$ of each map is given by  $[-1,1]$.
The coupling is only through the mean field
\begin{equation}
h_n={1\over N}\sum_{i=1}^Nf(x_{n}(i)) .
\end{equation}
There are a  great variety of phenomena generated by such globally coupled
maps,
and they seem to be due to two conflicting natures involved in the dynamics
\cite{kaneko1990a}.
On one side, the presence of a common
driving force coming from the mean field favors mutual synchronization.
On the other side, unstable growth of the difference in $x_n$
between any pair of elements tends to destroy such synchrony.
If the coupling constant $\epsilon$ is sufficiently large, mutual
synchronization
will be complete, all elements behaving identically. Then the collective motion
is
described simply by the single map $f(x)$.
On the contrary, if $\epsilon$ is sufficiently small,
the system will be ``turbulent'' where the elements
are scattered and behave chaotically in time.
In the present model there also appear band structures for intermediate
coupling strengths.
The state of perfect coherence loses stability at $a(1-e)=2$,
while the transition between turbulent and two-band states occurs at
$a(1-e)=\sqrt{2}$.
A phase diagram was obtained by Kaneko\cite{kaneko1992}.

In this paper, we shall confine our analysis to turbulent and two-band states.
The limit of large $N$ is of  our particular concern.
Some unexpected features of collective quantities such as the mean field have
been found recently.
For instance, the mean square deviation of such quantities dose
not tend to zero as $N\rightarrow\infty$ but saturates to a finite value.
This suggests the existence of nontrivial collective behavior
\cite{kaneko1992,just,pikovsky,kaneko1990b,kaneko1993,perez1993,perez1992,sinha}.

\section{Nonlinear Frobenius-Perron Equation}

For a very  large system size, direct numerical simulation of
(\ref{maps-defn}) requires enormous computer capacity.
Moreover, the occurrence of long transients may complicate
the analysis of stationary states and other asymptotic states.
For these reasons, an alternative approach seems desirable.

Statistical states of an ensemble of maps governed by (\ref{maps-defn}) can be
characterized by
its density distribution
\begin{equation}
\rho_n(x)={1\over N}\sum_{i=1}^N \delta(x-x_n(i)) ,
\label{density}
\end{equation}
where $\delta(x)$ is Dirac's delta function.
The evolution of $\rho_n(x)$ obeys the Frobenius-Perron equation
\cite{just,pikovsky}
\begin{equation}
\rho_{n+1}(x)=\int \delta(x-F_{n}(y))\rho_n(y)dy ,
\label{NFPE-int}
\end{equation}
where the effective map $F_n(x)$ depends on the mean field $h_n$ through
\begin{equation}
F_{n}(x)=(1-\epsilon)(1-a|x|)+\epsilon h_n .
\label{F}
\end{equation}
The mean field $h_n$ is calculated from the density (\ref{density}) as
\begin{equation}
h_n=\int f(x)\rho_n(x)dx .
\end{equation}
Equation (\ref{NFPE-int}) is called the nonlinear Frobenius-Perron equation
(NFPE) \cite{pikovsky}.
In the limit of large system size the density (\ref{density}) becomes a
``sufficient continuous'' function.
NFPE is completely analogous to the nonlinear Fokker-Planck equation.
In the following, we assume that the initial condition is given by a uniform
distribution
$\rho_0(x)={1\over2}$, that is, the elements of the original GCM are
distributed
in a completely random fashion.

Owing to the linearity of the two branches of the tent maps, (\ref{NFPE-int})
is simplified as
\begin{equation}
\rho_{n+1}(x)={\rho_n(y_+)+\rho_n(y_-)\over a(1-\epsilon)},
\label{NFPE}
\end{equation}
where $y_+$ and $y_-$ are the two preimages of the same $x$, i.e.\
$x=F_n(y_{\pm})$.

In the following, we propose an expansion form of NFPE in order to facilitate
the analysis of the model.
Figure \ref{fig:density} shows snapshots of $\rho_n(x)$ for two parameter
conditions.
The density distributions for a turbulent state and a two-band state are
displayed in Figs.\ref{fig:density}a
and \ref{fig:density}b, respectively.
We assume that the density distribution is a piecewise constant function,
which is due to the piecewise linearity of the tent map.
Then, one may simplify (\ref{NFPE}) by using a technique adopted
to analyze a one-dimensional tent map \cite{dorfle}.
The density distribution is expressed as
\begin{equation}
\rho_n(x)=\sum_{j=0}b_{n}^j \ \theta(c_{n}^j-x) ,
\label{rho}
\end{equation}
where $\theta(x)$ is a step function
\begin{equation}
\theta(x) = \left\{
\begin{array}{ll}
1 & \ (x\geq0) \\
0 & \ (x<0)
\end{array}
\right.
\end{equation}
In the above equation, $b_{n}^j$ and $c_{n}^j$ represent the height and the
location of
the jumps in $\rho_n(x)$, respectively.
Since the initial condition for the density distribution is assumed uniform,
$b_1^j$ and $c_1^j$ are given by
\begin{equation}
\left\{
\begin{array}{ccc}
b_1^0&=& \displaystyle {1\over a(1-\epsilon)} \\
c_1^0&=& \displaystyle 1-{a\epsilon\over2} \\
b_1^1&=& \displaystyle -{1\over a(1-\epsilon)} \\
c_1^1&=& \displaystyle 1-{a\epsilon\over2}-a(1-\epsilon) \\
others&=&0
\end{array}
\right.
\label{init}
\end{equation}
For $n=1$, the number of jumps in the density $\rho_n(x)$, which is denoted by
$\nu$, equals two, and
increases with $n$ as $\nu=n+1$.
With the use of (\ref{NFPE}), the evolutional equations for $b_{n}^j$ and
$c_{n}^j$
are given by
\begin{equation}
\left\{
\begin{array}{ccl}
b_{n+1}^0&=& \displaystyle {2\over a(1-\epsilon)}\sum_{j'=0} b_{n}^{j'} \
\theta(c_{n}^{j'}) \\
c_{n+1}^0&=& \displaystyle F_n(0) \\
b_{n+1}^{j+1}&=& \displaystyle -{1\over a(1-\epsilon)}\, b_{n}^j \,(2
\theta(c_{n}^j)-1) \\
c_{n+1}^{j+1}&=& \displaystyle F_n(c_{n}^j)
\end{array}
\right.
\label{expansion1}
\end{equation}
This set of equations gives an alternative representation of NFPE (\ref{NFPE}).
Note the identity
\begin{equation}
\sum_{j=0}b_n^j=0
\end{equation}
which results from the requirement $\rho_n(-1)=0$.
By using (\ref{density}) along with (\ref{rho}), the mean field $h_n$ is
calculated as
\begin{equation}
h_n=(1-{a\over2}\sum_{j=0}b_{n}^j \, c_{n}^j \, |c_{n}^j|) .
\label{expansion2}
\end{equation}
Equations (\ref{F}), (\ref{expansion1}) and (\ref{expansion2}) constitute a
closed set of
exact nonlinear equations equivalent to NFPE,
and (\ref{init}) specifies the initial condition.

The dimension of the dynamical system (\ref{expansion1}) is potentially
infinite ($j=0,1,\dots ,\infty$)
because the number of jumps in $\rho_n(x)$ increases indefinitely with $n$.
We now approximate this by a finite number.
We expect that the jumps of sufficiently large $j$ hardly contribute to the
collective behavior,
so that we may ignore such minor jumps.
Indeed, the third equation in (\ref{expansion1}) means that  the height of the
jumps is given by
\begin{equation}
|b_{n}^{j}|={1\over a^j(1-\epsilon)^j}|b_{n-j}^0| .
\end{equation}
Thus, $b_{n}^j$ approaches 0 exponentially as $j\rightarrow\infty$, provided
$a(1-\epsilon)>1$. We now focus on this incoherent regime .
Note that in the opposite case, i.e.\ $a(1-\epsilon)<1$, $\rho_n(x)$ becomes a
delta function
since all elements are in perfect synchrony, so that the breakdown of the above
approximation is apparent.
Our approximate method seems more convenient than those employed previously
\cite{kaneko1992,pikovsky}.
In what follows, we restrict the maximum number of the jumps to 50
\cite{coment2}.
The mean field behavior calculated with this approximation turns out
consistent with our direct numerical simulation.

\section{Stability of stationary states}

We investigate stationary solutions i.e.\ fixed points of NFPE which are
however not the fixed points
of GCM (\ref{maps-defn}).
The fixed points of NFPE are determined by
$\rho_{n+1}(x)=\rho_{n}(x)\equiv\rho_*(x)$.
{}From the expansion (\ref{expansion1}) in the last section, this condition is
given by
\begin{equation}
\left.
\begin{array}{lcl}
b_{*}^0&=& \displaystyle {2\over a(1-\epsilon)}\sum_{j'=0} b_{*}^{j'} \
\theta(c_{*}^{j'}) , \\
c_{*}^0&=& \displaystyle F_*(0) ,\\
b_{*}^{j+1}&=& \displaystyle -{1\over a(1-\epsilon)}\,b_{*}^j\, (2
\theta(c_{*}^j)-1) ,\\
c_{*}^{j+1}&=& \displaystyle F_*(c_{*}^j) ,
\end{array}
\right.
\label{fixed}
\end{equation}
where the map $F_*$ evaluated at the fixed point $\rho_*$ is given as
\begin{eqnarray}
F_*(x)&=&(1-\epsilon)(1-a|x|)+\epsilon h_* ,
\label{F2} \\
h_*&=&(1-{a\over2}\sum_{j=0}b_{*}^j \, c_{*}^j \, |c_{*}^j|) .
\label{fixed2}
\end{eqnarray}
The above equations constitute a set of self-consistent equations.
For the time-being, we regard $h_*$ in (\ref{F2}) as an input field $h_{in}$,
and distinguish it from $h_*$ in (\ref{fixed2}) which we regard as an output
field $h_{out}$.
We can easily calculate $b_{*}^j$ and $c_{*}^j$
from (\ref{fixed}) and the normalization condition for the density
\begin{equation}
\sum_{j=0}b_*^{j}(c_*^{j}+1)= 1 .
\end{equation}
With $b_{*}^j$ and $c_{*}^j$ thus obtained, the static field $h_{out}$ is given
by (\ref{fixed2}).
Consequently, $h_{out}$ is obtained as a function of $h_{in}$, i.e.\
$h_{out}(h_{in};a,e)$, which is called a ``static mapping'' \cite{perez1992}.
The fixed points of this mapping give the stationary solutions of our system.
The mapping actually has some fixed points in most of the parameter region
under consideration.

Numerical stability analysis of the stationary states is now presented.
We first linearize (\ref{expansion1}) about a stationary state ($b_{*}^j$,
$c_{*}^j$)
corresponding to the density $\rho_{*}(x)$, and
consider a small density deviation from it.
Noting that $F_n(x)$ in (\ref{expansion1}) depends on $b_{n}^j$ and $c_{n}^j$,
we have a set of linear maps given by
\begin{equation}
\left\{
\begin{array}{lcl}
\beta_{n+1}^0&=& \displaystyle
{2\over a(1-\epsilon)}\sum_{j'=0} \beta_{n}^{j'} \ \theta(c_{*}^{j'}) \\
\gamma_{n+1}^0&=& \displaystyle
-{a\epsilon \over2}\sum_{j'=0} \{ \beta_{n}^{j'} \, c_{*}^{j'} \, |c_{*}^{j'}|
+
2\gamma_{n}^{j'} \, b_*^{j'}\, |c_{*}^{j'}| \} \\
\beta_{n+1}^{j+1}&=& \displaystyle
-{1\over a(1-\epsilon)}\, \beta_{n}^j \, (2 \theta(c_{*}^j)-1) \\
\gamma_{n+1}^{j+1}&=& \displaystyle
-{a\epsilon \over2}\sum_{j'=0} \{ \, \beta_{n}^{j'} \, c_{*}^{j'} \,
|c_{*}^{j'}| +
2\gamma_{n}^{j'} \, b_*^{j'} \, |c_{*}^{j'}| \}
- a(1-\epsilon)\, \gamma_{n}^j \, (2 \theta(c_*^{j})-1)
\end{array}
\right.
\label{matrix}
\end{equation}
where $\beta_{n}^j=b_{n}^j-b_{*}^j$ and  $\gamma_{n}^j=c_{n}^j-c_{*}^j$.
The set of equations above can be expressed with a coefficient matrix $D$ as
\begin{equation}
\left(\begin{array}{l}
\vec{\beta}_{n+1} \\ \vec{\gamma}_{n+1}
\end{array}\right)
=D \
\left(\begin{array}{l}
\vec{\beta}_{n}\\ \vec{\gamma}_{n}
\end{array}\right) .
\end{equation}
The eigenvalues of $D$ determines the stability of the stationary solution.
It is obvious from (\ref{matrix}) that $D$ has the form
\begin{equation}
D=
\left(\begin{array}{cc}
D_1 & 0 \\
\ast & D_2
\end{array}\right) .
\end{equation}
Thus, the eigenvalues of $D$ are given by those of $D_1$ and $D_2$,
and they are generally complex numbers.
In Fig.\ref{fig:eigen0}, we show the eigenvalues of $D$ for $a=1.5$ and
$\epsilon=0.05$
for which a stationary state exists stably.
They include real numbers and complex-conjugate pairs, and their distribution
is roughly given
by two concentric circles.
The inner and outer circles correspond to $D_1$ and $D_2$, respectively.
The normalization condition for $\rho_n(x)$ requires that one of the
eigenvalues is always unity \cite{just}.
If all of the remaining eigenvalues lie within the unit circle, the stationary
solution
remains stable. It loses stability in the following two ways.

The first case is due to band splitting.
This occurs at $a(1-\epsilon)=\sqrt{2}$ \cite{coment}.
In the unstable region, period-two motion appears, and then
the density $\rho_n(x)$ has a two-band structure as shown in
Fig.\ref{fig:density}b.
An infinite number of period-two solutions exist and a particular state is
selected out
by imposing a specific initial condition. Therefore, it is expected that their
stability is marginal.

Figure \ref{fig:eigen}a shows the eigenvalues under the condition that
stationary state is stable ($a(1-\epsilon)<\sqrt{2}$).
The moduli of all eigenvalues except for one with vanishing argument are
smaller than unity.
Figure \ref{fig:eigen}b shows the eigenvalues after band splitting
($a(1-\epsilon)>\sqrt{2}$).
The matrix $D_1$ has two eigenvalues with moduli 1, while
their arguments are given by to $0$ and $\pi$ (the eigenvalues are $1$ and
$-1$, respectively).
In Fig.\ref{fig:eigen2}, we plotted the eigenvalue $\lambda_\pi$ with argument
$\pi$ as a function of $\epsilon$.
As is seen from Fig.\ref{fig:eigen2}a, $|\lambda_\pi|$ approaches unity as
$\epsilon$ increases up to the band-splitting point $\epsilon_0$ under fixed
$a$ at $1.5$.
For this value of $a$, the band splitting is seen for
$\epsilon>\epsilon_0={3-2\sqrt{2}\over3}=0.05719\cdots$,
and $|\lambda_\pi|$ is kept at unity over this range of $\epsilon$.
The results of the  stability analysis agree with our observation from direct
numerical simulation.
In order to investigate the nature of this instability in some detail,
we make a log-log plot near the band-splitting point. This is shown in
Fig.\ref{fig:eigen2}b.
As is seen from Fig.\ref{fig:eigen2}b, $|\lambda_\pi|$ approaches unity
linearly with $\epsilon_0-\epsilon$.
Thus, the band-splitting point can be determined accurately from this kind of
analysis.

There is another type of instability.
This is associated with a Hopf bifurcation, and occurs even in the region where
the band splitting
is not seen ($a(1-\epsilon)<\sqrt{2}$).
Figure \ref{fig:eigen}c shows the distribution of the eigenvalues in the
unstable regime.
$|\lambda_\pi|$ is smaller than unity and
the matrix $D_2$ has an eigenvalue with modulus larger than unity.
The argument of this eigenvalue is about $0.86\pi$.
In this region, the collective behavior of the original system is found
quasiperiodic.
The evolution of the corresponding mean field $h_n$ is displayed in
Fig.\ref{fig:h},
and its power spectrum is given in Fig.\ref{fig:power}.
The latter  has some sharp peaks, and the frequency of the highest peak is
about $0.43$.
This value is consistent with the results of our stability analysis.

\section{summary}
The piecewise-linearity of the map allows us to make an expansion of NFPE
in term of the height and location of the jumps in the density distribution.
We analyzed the stability of stationary solutions by using this expansion.
Linearization of the expansion is expressed with two matrices $D_1$ and $D_2$.
The band-splitting instability is associated with the eigenvalues of $D_1$.
Stability analysis predicts the band-splitting point accurately.
The band splitting results in marginally stable period-two states.
There is also a region where quasi-periodic motions appear.
The corresponding instability is essentially a Hopf bifurcation, and associated
with
the eigenvalues of $D_2$.

Our analysis presupposes infinite system size.
In actual finite system, however, we expect some fluctuations around the
solutions of NFPE.

We investigated only globally coupled tent maps.
It should be mentioned, however, that the same methods could also be useful for
investigating
other piecewise linear maps.

\acknowledgements{
The author would like to acknowledge valuable advices from Y.Kuramoto
and fruitful discussions with T.Chawanya and the members of Nonlinear
Dynamics group of Kyoto University.
}

\begin{figure}
\noindent
\caption{
Snapshot of the density distribution $\rho_n(x)$ after transients have
subsided.
(a) Turbulent state obtained from the expansion of NPFE (see Section 2);
$a=1.6$, $\epsilon=0.05$.
(b) Two-band state; $a=1.6$, $\epsilon=0.15$.
}
\label{fig:density}
\end{figure}

\begin{figure}
\noindent
\caption{
Eigenvalues of the $50 \times 50$ matrices $D_1$ and $D_2$ for $a=1.5$ and
$\epsilon=0.05$ where
the stationary solution remains stable.
}
\label{fig:eigen0}
\end{figure}

\begin{figure}
\noindent
\caption{
Eigenvalues of the matrices $D_1$ and $D_2$.
Their absolute values vs.\ arguments are shown. Since the eigenvalues appear as
complex-conjugate pairs,
only those with arguments lying between 0 and 
(a) $a=1.5$, $\epsilon=0.05$. The moduli of all eigenvalues are smaller than 1
except one with vanishing argument.
(b) $a=1.5$, $\epsilon=0.06$. There exists an eigenvalue of modulus 1 and
argument $\pi$.
(c) $a=1.9$, $\epsilon=0.12$. There exists an eigenvalue with modulus is larger
than 1.
}
\label{fig:eigen}
\end{figure}

\begin{figure}
\noindent
\caption{
(a) Eigenvalue $|\lambda_\pi|$ with argument $\pi$ vs.\ $\epsilon$
where $a=1.5$. (b) Log-log plot near the band-splitting point $\epsilon_0$
where $a=1.5$.
}
\label{fig:eigen2}
\end{figure}

\begin{figure}
\noindent
\caption{
Trajectory of the mean field $h_n$ obtained from the expansion of NPFE (see
section 2).
The parameter values are the same as in Fig.2c (i.e.\ $a=1.9$,
$\epsilon=0.12$).
This was obtained from 1000 steps after transients have subsided.
}
\label{fig:h}
\end{figure}

\begin{figure}
\noindent
\caption{
Power spectrum of the time series of the mean field $h_n$ obtained from the
expansion
of NPFE (see section 2).
This was obtained from an average over 10 runs each of 1024 iterations after
transients have subsided.
Parameter values are the same as in Fig.2c.
}
\label{fig:power}
\end{figure}


\begin{references}
\bibitem{coment2}
The parameters we used satisfy $a(1-\epsilon)\simeq\sqrt{2}$.
Therefore the height of the jumps ignored is not larger than  $2^{-25}$.
\bibitem{coment}
Higher order band splitting is also seen.
At the critical parameter values $a(1-\epsilon)=2^{1/2^n}$, $2^n$-band
structure appears.
The period of this state is $2^n$. If the population in some bands is zero,
the number of bands can be  smaller than $2^n$.
\end{references}
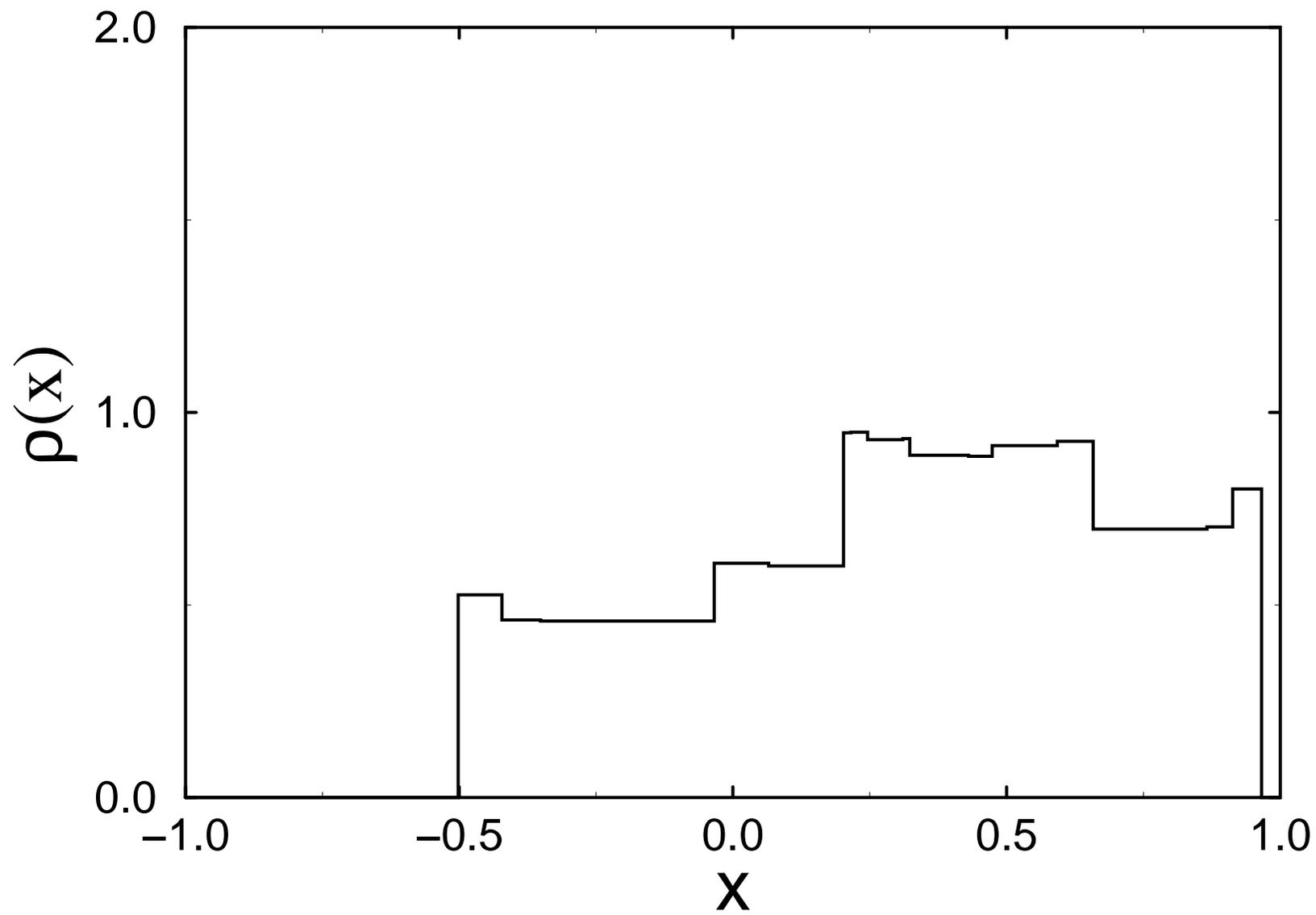
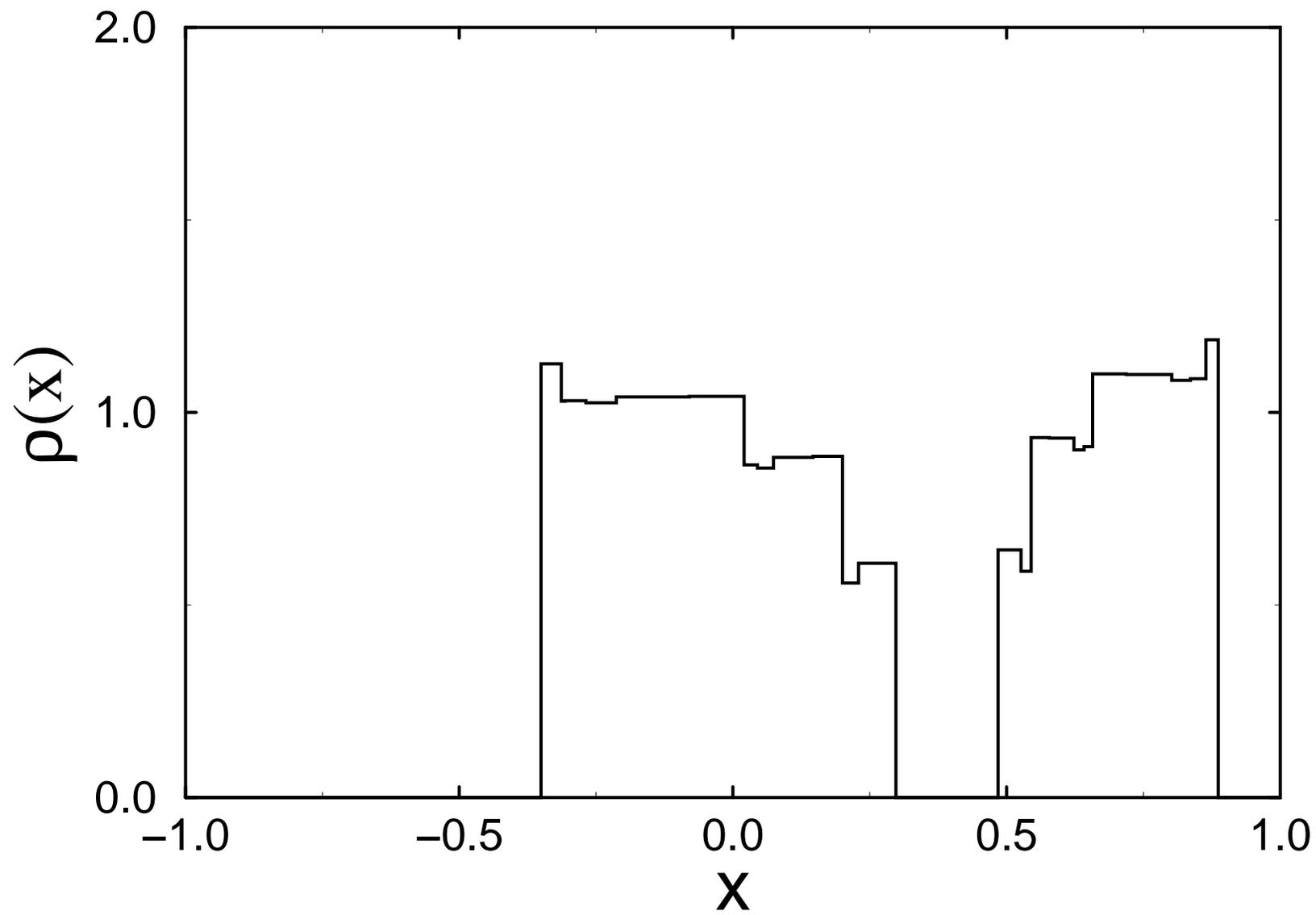
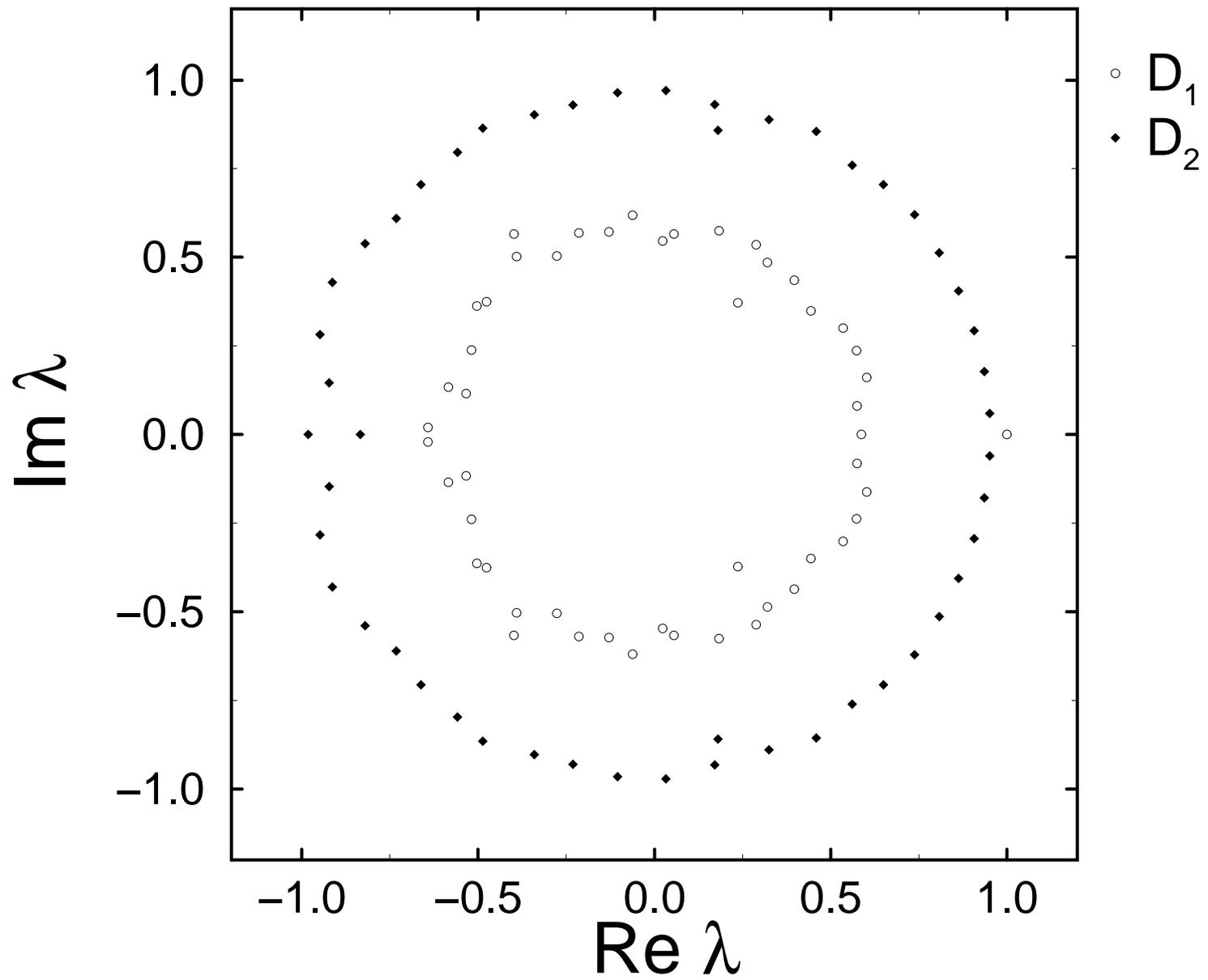
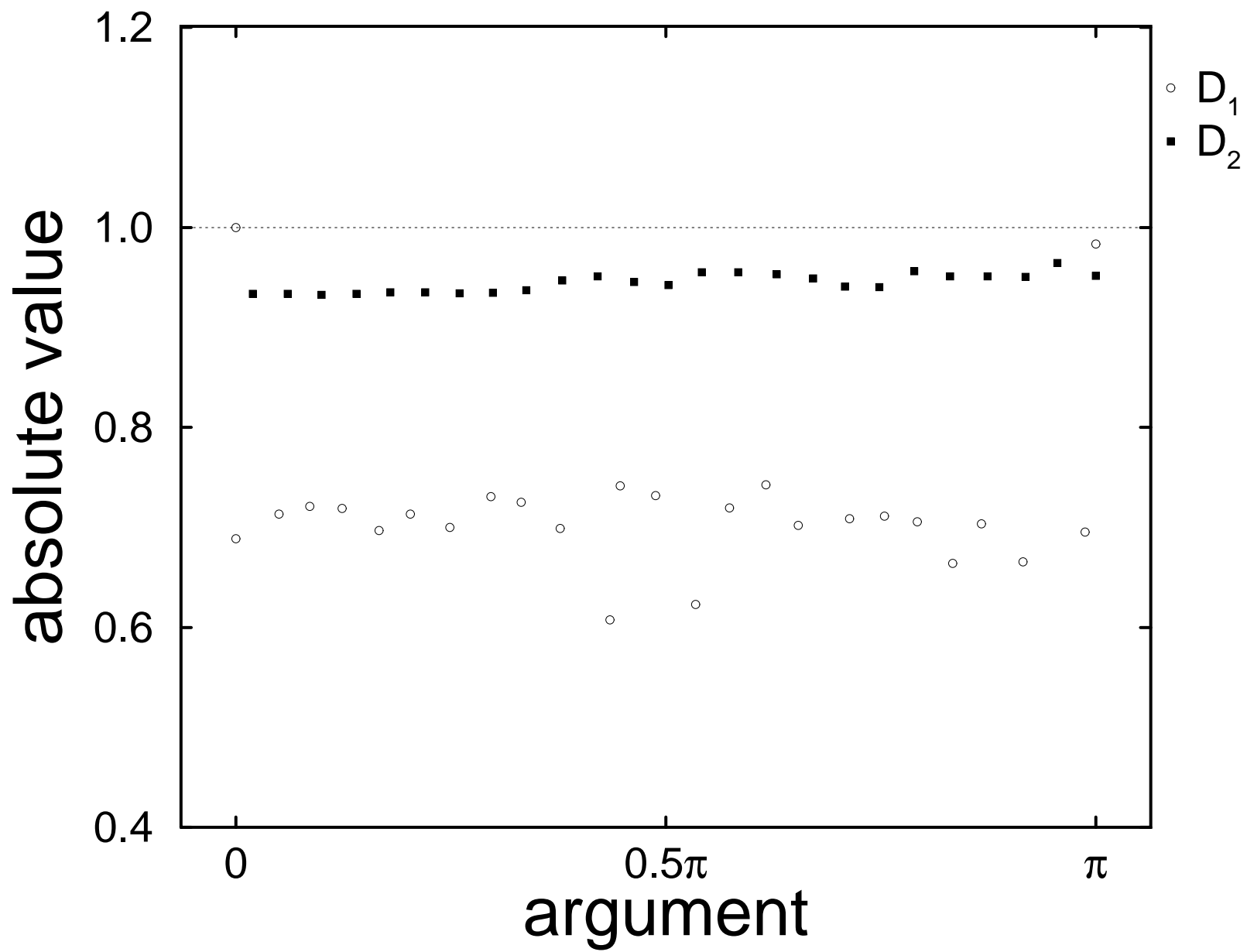
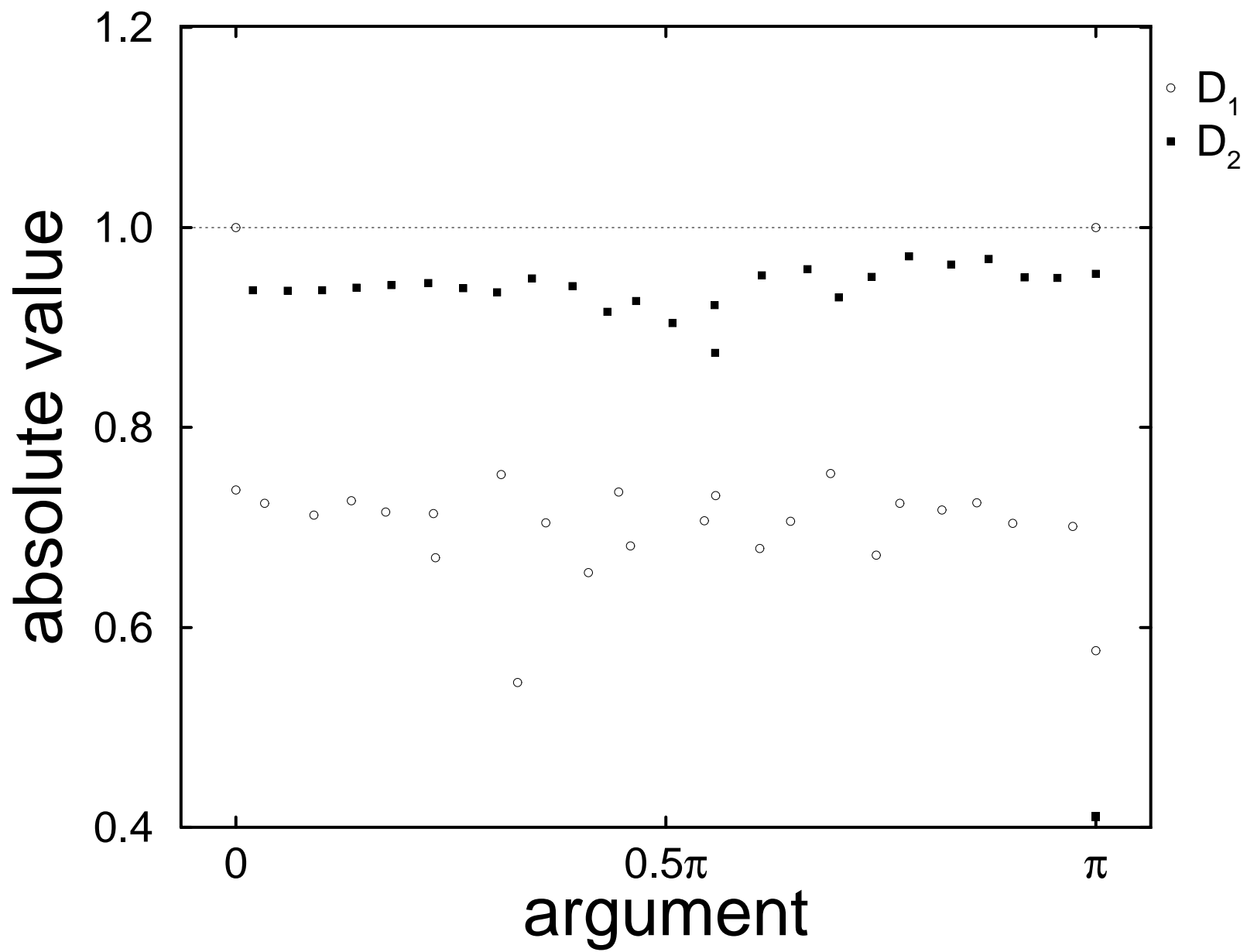
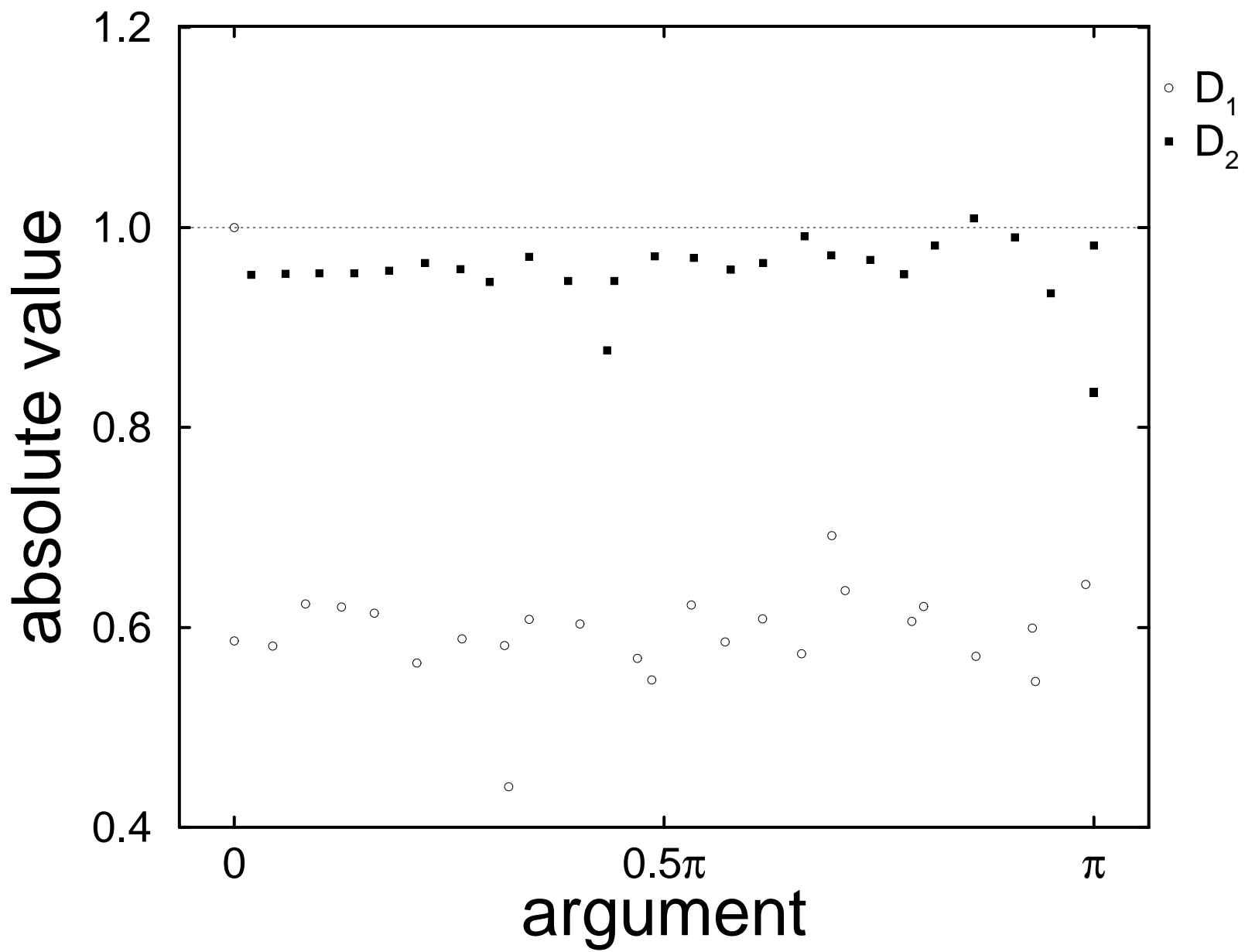
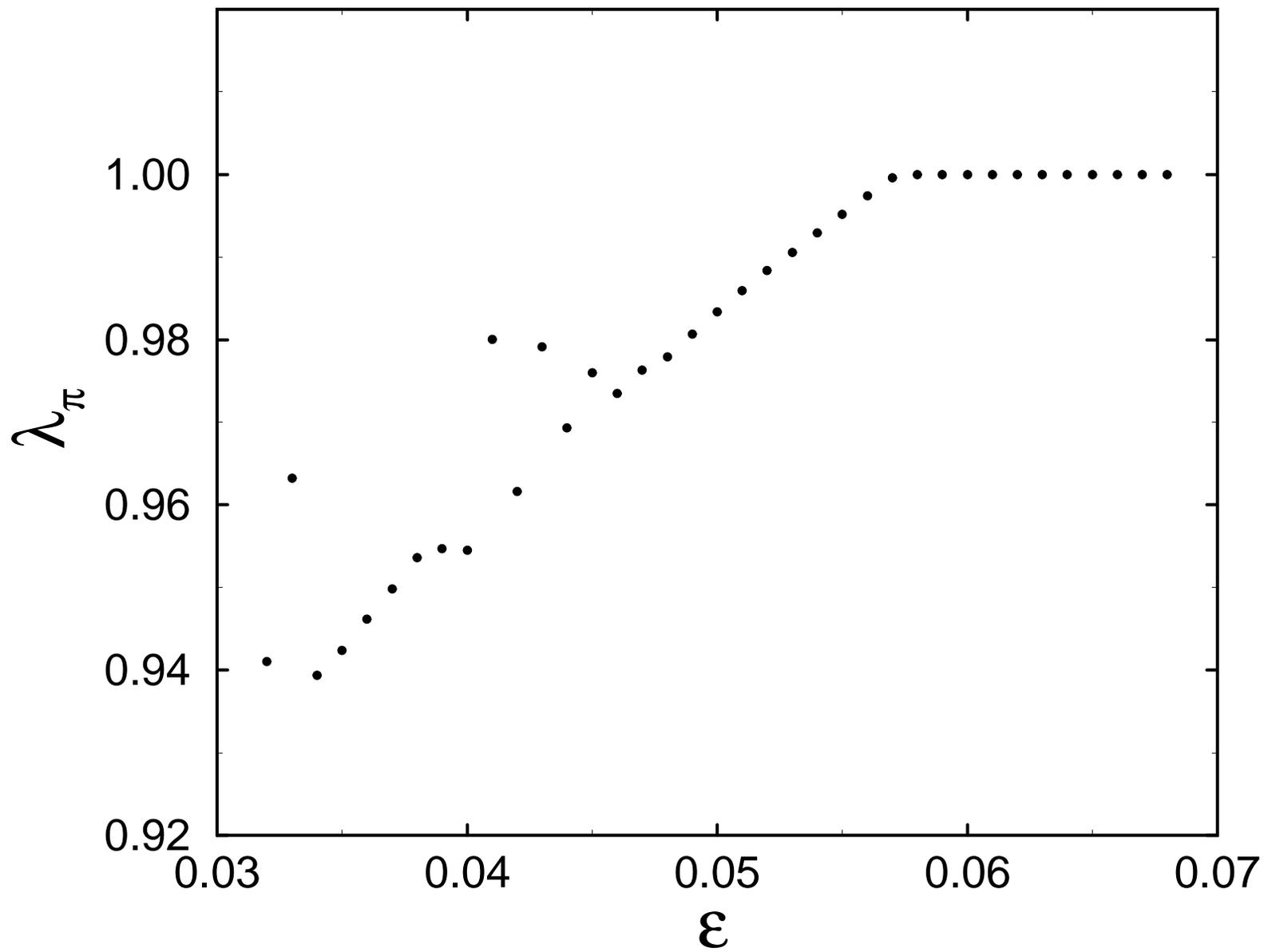
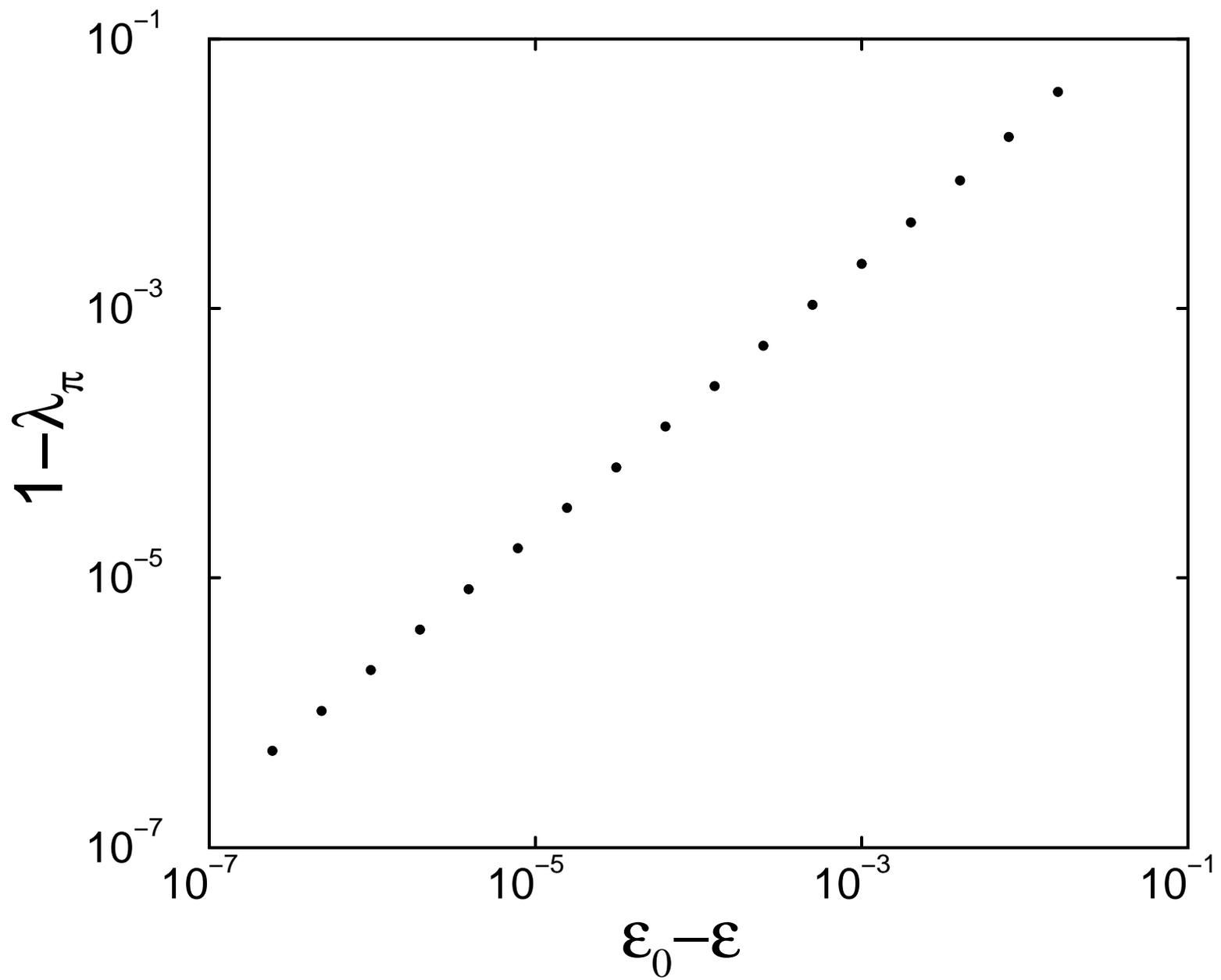
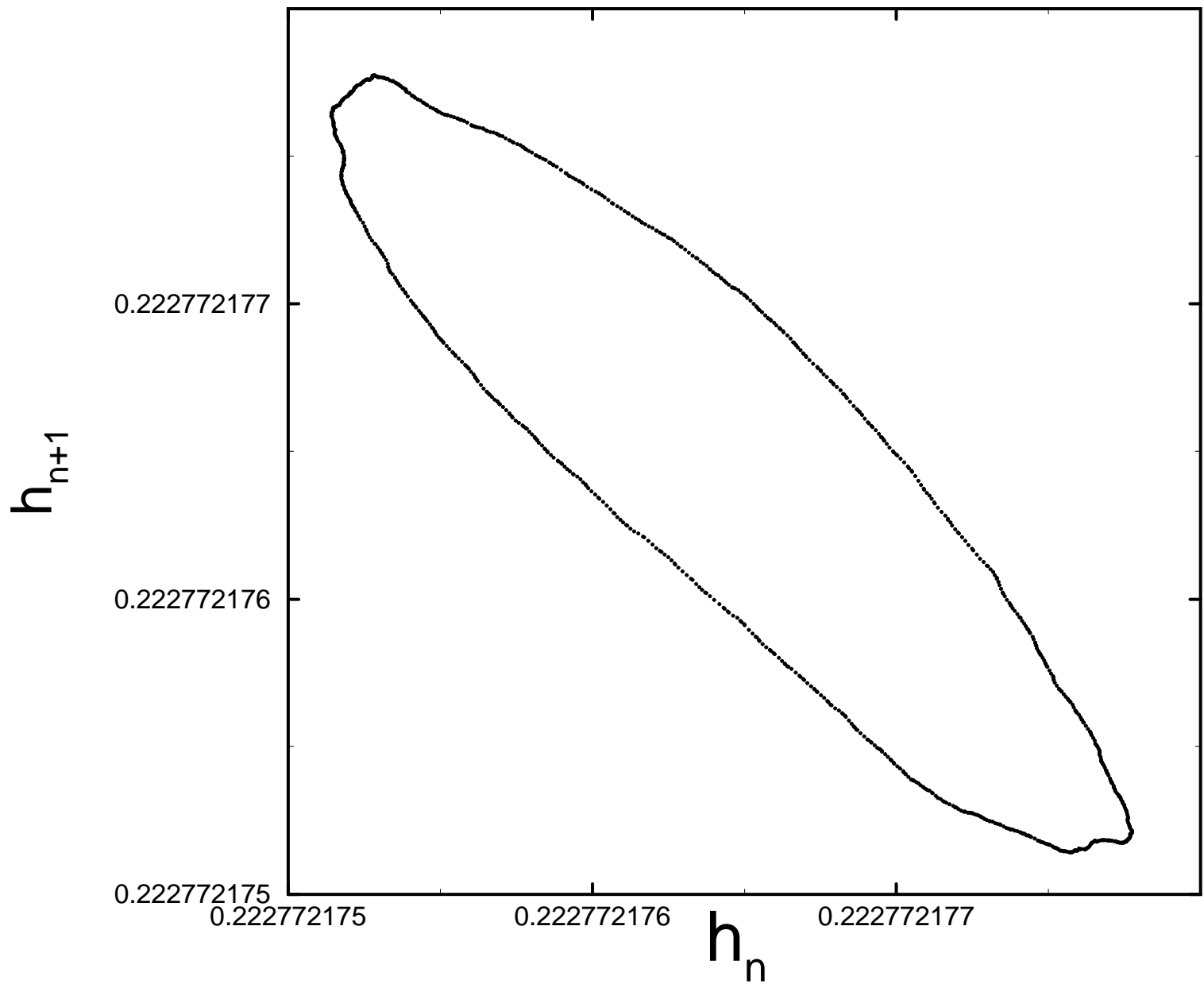
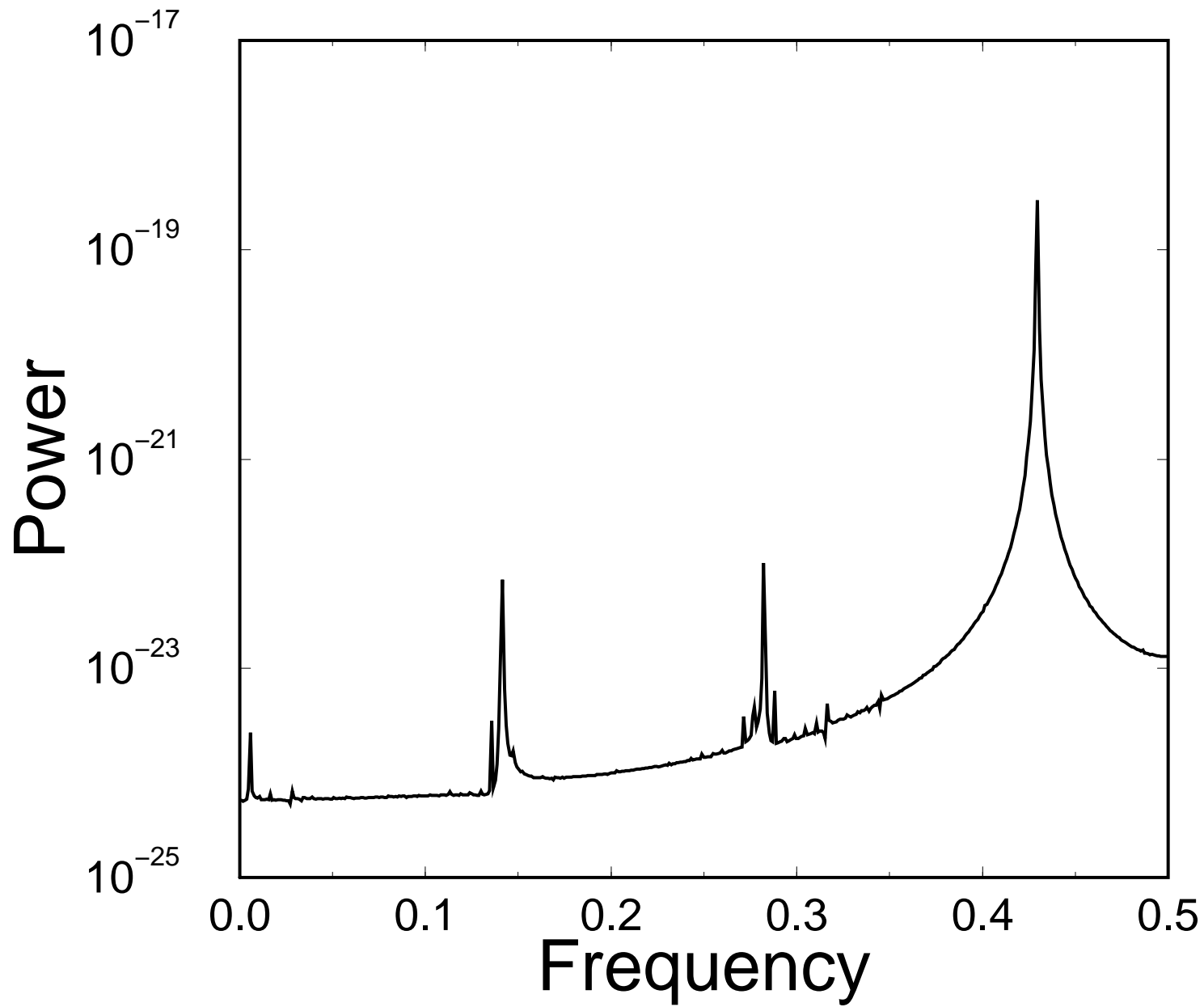
\end{document}